\documentclass{ieeeaccess}
\usepackage{cite}
\usepackage{amsmath,amssymb,amsfonts}
\usepackage{algorithm}
\usepackage{algpseudocode}
\usepackage{etoolbox}
\usepackage{graphicx}
\usepackage{textcomp}
\usepackage{mathtools}
\usepackage{amsmath}
\usepackage{graphicx}
\usepackage{caption}
\usepackage{url}

\usepackage[table]{xcolor}

\AtBeginEnvironment{algorithm}{\color{black}}

\definecolor{low}{RGB}{255,255,204} 
\definecolor{high}{RGB}{154,205,50}    
\definecolor{low}{RGB}{255,255,255}  
\definecolor{high}{RGB}{179,213,149}   
\newcommand{\gradient}[1]{\cellcolor{high!\the\numexpr#1\relax!low!100}}

\def\BibTeX{{\rm B\kern-.05em{\sc i\kern-.025em b}\kern-.08em
    T\kern-.1667em\lower.7ex\hbox{E}\kern-.125emX}}
\begin{document}
\history{Date of publication xxxx 00, 0000, date of current version xxxx 00, 0000.}
\doi{10.1109/ACCESS.2017.DOI}

\title{Bridging the Question-Answer Gap in Retrieval-Augmented Generation:\\
       Hypothetical Prompt Embeddings}
       
\author{
    \uppercase{Domen Vake}\authorrefmark{1,2},
    \MakeUppercase{Jernej Vičič}\authorrefmark{1,3}, and 
    \MakeUppercase{Aleksandar Tošić}.\authorrefmark{1,2},
}

\address[1]{University of Primorska, Faculty of Mathematics, Natural Sciences and Information Technologies}
\address[2]{InnoRenew CoE}
\address[3]{Research Centre of the Slovenian Academy of Sciences and Arts, The Fran Ramovš Institute}

\footnote{This project received funding from the European Union’s Horizon 2020 grant \#101135012.}

\markboth
{Vake \headeretal: Bridging the Question-Answer Gap in Retrieval-Augmented Generation: Hypothetical Prompt Embeddings}
{Vake \headeretal: Bridging the Question-Answer Gap in Retrieval-Augmented Generation: Hypothetical Prompt Embeddings}

\corresp{Corresponding author: Domen Vake (e-mail: domen.vake@famnit.upr.si).}
\begin{abstract}
Retrieval-Augmented Generation (RAG) systems synergize retrieval mechanisms with generative language models to enhance the accuracy and relevance of responses. However, bridging the style gap between user queries and relevant information in document text remains a persistent challenge in retrieval-augmented systems, often addressed by runtime solutions (e.g., Hypothetical Document Embeddings (HyDE)) that attempt to improve alignment but introduce extra computational overhead at query time. To address these challenges, we propose Hypothetical Prompt Embeddings (HyPE), a framework that shifts the generation of hypothetical content from query time to the indexing phase. By precomputing multiple hypothetical prompts for each data chunk and embedding the chunk in place of the prompt, HyPE transforms retrieval into a question-question matching task, bypassing the need for runtime synthetic answer generation. This approach does not introduce latency but also strengthens the alignment between queries and relevant context. Our experimental results on six common datasets show that HyPE can improve retrieval context precision by up to 42 percentage points and claim recall by up to 45 percentage points, compared to standard approaches, while remaining compatible with re-ranking, multi-vector retrieval, query decomposition, and other RAG advancements.
\end{abstract}


\titlepgskip=-15pt

\maketitle

\section{Introduction}
\label{sec:introduction}
\PARstart{R}{}etrieval-augmented generation 
 (RAG) systems have emerged as a powerful paradigm in natural language processing, combining the strengths of retrieval-based approaches with the generative capabilities of large language models (LLMs)~\cite{lewis2020retrieval}. By leveraging external knowledge sources, RAG systems enhance the factual accuracy and relevance of generated responses, addressing the limitations of standalone generative models, such as outdated knowledge or limited access to restricted information. Despite their success, existing RAG implementations often struggle with aligning retrieval and generation in a way that efficiently bridges the gap between user queries and relevant document content. 

To optimize retrieval performance, researchers have explored a variety of strategies, ranging from efficient chunking (splitting texts into coherent subunits)\cite{lewis2020retrieval, yepes2024financial} to re-ranking methods that refine initial retrieval results via cross-encoders or boosted similarity scoring\cite{mishra2024searchd}. Advanced frameworks such as GraphRAG exploit graph structures to capture cross-document relationships for more nuanced multi-hop or contextual queries~\cite{yasunaga2021qa}. Meanwhile, domain adaptation techniques focus on tailoring retrieval to specialized topics, ensuring that queries can be matched effectively, even in areas where language models might otherwise lack expertise.

Despite these efforts, a persistent hurdle in RAG remains the mismatch between user queries, which typically adopt an interrogative style, and corpus content, which is usually expository or declarative in nature. This style difference hampers the alignment of query embeddings with document embeddings, occasionally allowing key information to go unretrieved. A notable solution to this problem is Hypothetical Document Embeddings (HyDE)~\cite{gao2022precise}, which prompts an LLM at query time to generate a synthetic answer, then uses that short text as the query for retrieval. 

In this paper, we introduce Hypothetical Prompt Embeddings (HyPE), a new approach that tackles query–document style mismatch without adding overhead to every user request. Rather than generating synthetic answers at inference, HyPE precomputes multiple hypothetical questions for each corpus chunk at indexing time. These question-like prompts are embedded and stored, so that query matching effectively becomes a question–question retrieval problem. By shifting hypothetical generation offline, HyPE avoids additional runtime LLM calls. 

To evaluate HyPE's effectiveness, we compare it against a naive RAG implementation and HyDE across multiple datasets and evaluation metrics, including precision, recall, and faithfulness. Our results demonstrate that HyPE offers substantial improvements in retrieval efficiency, reducing the computational burden while achieving comparable or better retrieval accuracy and contextual relevance.

The contributions of this paper are as follows:
\begin{itemize}
    \item We introduce the concept of precomputed hypothetical prompt embeddings to optimize retrieval efficiency in RAG systems.
    \item We provide a comprehensive performance comparison between a Naive RAG implementation, HyDE, and HyPE.
    \item We present experimental results showcasing the trade-offs in retrieval quality and the effect of retrieval approach on generation across various datasets.
\end{itemize}

By shifting the hypothetical generation process from runtime to indexing, HyPE represents a scalable and efficient alternative for RAG systems, offering practical benefits for real-world applications requiring fast and reliable retrieval-augmented text generation.
The rest of the paper is structured as follows: Section \ref{related} presents the related works and surveys, Section \ref{methodology} presents the used methodology. Follows the presentation of the experiment setting with the presentation of the datasets and the evaluation metrics. Section~\ref{analysis} presents the results with thorough analysis, Section \ref{conclusion} presents the finals conclusions.

\section{Related Works}
\label{related}

Lewis et al.~\cite{lewis2020retrieval} introduced the original RAG framework, combining dense retrieval with generation to mitigate hallucination and provide grounding via external documents. While effective, early RAG pipelines~\cite{lewis2020retrieval, izacard2022few} often suffer from limitations in retrieval precision by simply embedding the user queries and retrieving text chunks via approximate nearest neighbour (ANN) search over a vector index. However, a persistent challenge remains: user queries are often phrased in question form, whereas documents or chunks are stored in an expository or statement-oriented style, creating a semantic or “lexical–conceptual” gap~\cite{furnas1987vocabulary, nogueira2019document}. This mismatch can degrade the retrieval’s accuracy and ultimately weaken the generative model’s faithfulness.

 Cross-encoder reranking and multi-vector representations~\cite{xiong2021ance} enhance retrieval quality post hoc, but introduce additional inference-time latency.
 
One direction of research attempts to alleviate this mismatch by expanding documents with likely queries. For instance, Doc2Query~\cite{nogueira2019doc2query} uses a sequence-to-sequence model (often T5) to generate synthetic questions for each document, appending them to the text so that a bag-of-words ranker like BM25~\cite{robertson2009probabilistic} can match real user queries more easily. While Doc2Query often boosts first-stage recall, subsequent studies noted that the generation process can hallucinate irrelevant expansions, leading to index bloat. Among them, Doc2Query-- (Minus-Minus)~\cite{gospodinov2023doc2query} addresses this by filtering out low-quality expansions, improving accuracy and reducing index size for BM25-based retrieval. However, these methods predominantly operate in a lexical retrieval space rather than dense embeddings, and they store expansions as text appended to each document and, as such, act more like an enrichment of the chunks.

Another related direction focuses on generating synthetic training data to train or fine-tune dense retrievers in new domains. Ma et al.~\cite{ma2020zero} propose using a question-generation model, trained on general-purpose Question-Answer(QA) pairs, to generate “pseudo-queries” for domain-specific passages. A dual-encoder retrieval model is then trained on these synthetic (question, passage) pairs—effectively learning domain adaptation in a zero-shot setting. While powerful for building domain-specialized retrievers, the method requires re-training or fine-tuning a dense model on large-scale synthetic data. By contrast, our approach bypasses model re-training at retrieval time and, instead, alters how we store the passages (i.e., their hypothetical question embeddings).

More recently, HyDE~\cite{gao2022precise} addresses query–document mismatch by generating a hypothetical answer or short passage at query time. Instead of embedding the user’s question directly, HyDE prompts an LLM to produce an approximate response, then embeds that synthetic text. This is used to retrieve relevant real documents from a vector index. While HyDE can improve retrieval accuracy for zero-shot question answering, it incurs an extra inference cost per user query. Additionally, the method may struggle, where the prompt queries for niche domain knowledge, where the model may not have sufficient knowledge to produce a representative sample.
Building on this line of work, Eibich et al.~\cite{eibich2024aragog} conducted an empirical study comparing RAG retrieval enhancements, including HyDE, reranking, multi-query expansion, and maximal marginal relevance (MMR). Their findings highlight HyDE's strong performance in both recall and faithfulness metrics, while also noting its trade-offs in runtime efficiency. 

Several surveys have recently addressed retrieval-augmented systems. Gupta et al.\cite{gupta2024survey} and Cheng et al.\cite{cheng2025knowledge} offer extensive taxonomies of retrieval mechanisms, identifying open challenges in domain adaptation, embedding alignment, and inference efficiency. Notably, both surveys identify the need for methods that preserve retrieval quality while reducing reliance on runtime LLM calls.

\begin{figure*}[ht]
\centering
\begin{minipage}{1\textwidth}
  \centering
  \includegraphics[width=1\linewidth]{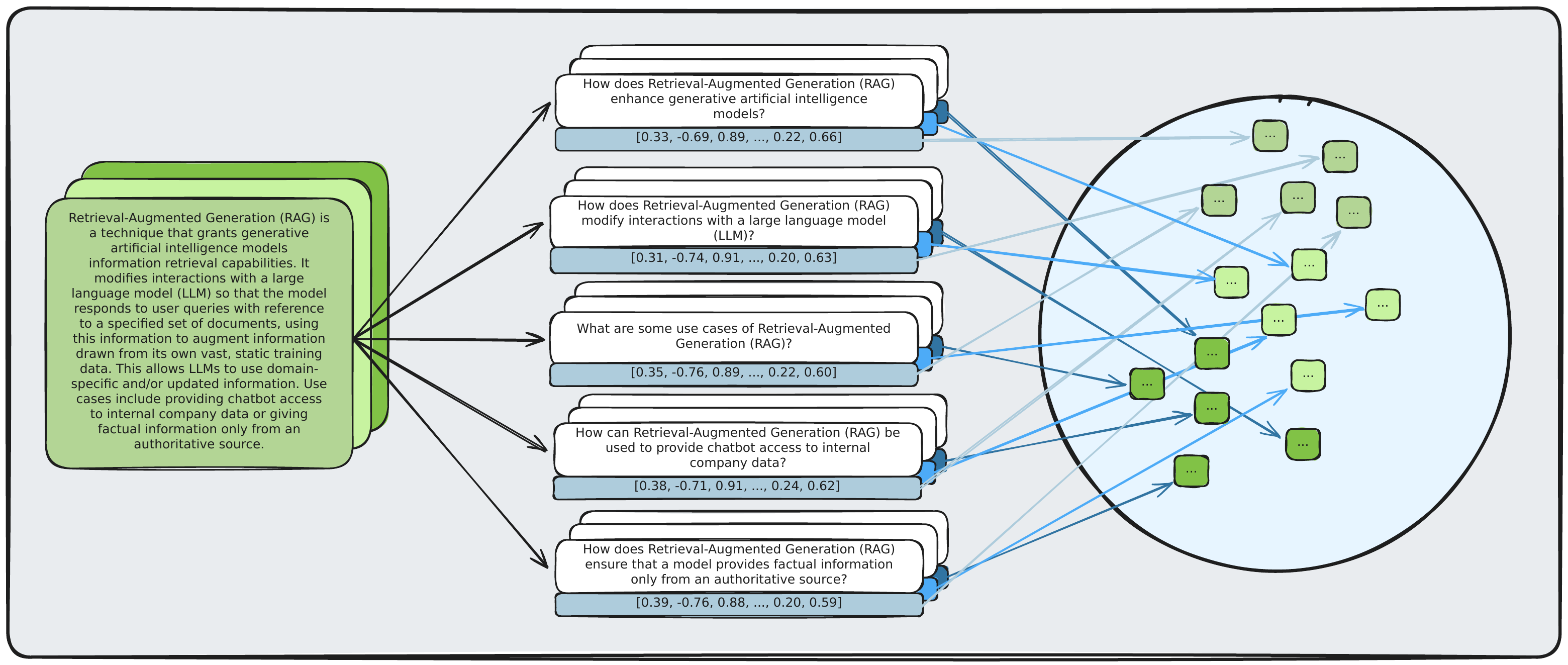}
  \caption{Illustration of the Hypothetical Prompt Embeddings (HyPE) framework, showcasing the process of precomputing hypothetical questions during indexing to optimize retrieval efficiency in Retrieval-Augmented Generation (RAG) systems.}
  \label{fig:test1}
\end{minipage}%
\end{figure*}

In this context, our proposed Hypothetical Prompt Embeddings (HyPE) introduces a novel retrieval strategy that pre-computes hypothetical question-style prompts at indexing time. This shifts the burden of LLM generation to the offline phase.

\section{Methodology}
\label{methodology}
HyPE addresses the challenge of aligning user queries and relevant content by pre-computing hypothetical prompts at the indexing stage, contrasting with HyDE’s runtime generation of synthetic answers. This shift avoids additional inference overhead per query and improves retrieval precision by ensuring that both user queries and stored embeddings share a question-like form.

The method begins by splitting the corpus $D$ into coherent chunks \newline$C_1, C_2, …, C_n$, where each chunk provides a self-contained unit of information. For each chunk $C_i$, an LLM $G$ generates multiple hypothetical prompts $Q_i={q_{i1},q_{i2},…,q_{ik}}$, simulating possible user queries that the chunk might answer. This offline step does not introduce any additional computational cost at query time, as no new prompts need to be generated for each user request.

Each hypothetical prompt $q_{ij}$ is then mapped to an embedding \newline$v_{ij}=f(q_{ij})\in\mathbb{R}^d$ using a pre-trained dense retrieval model $f$. Rather than storing these prompt embeddings separately, we associate each $v_{ij}$ with the original chunk $C_i$, thus building an index of vector–chunk pairs:

\begin{gather*}
E=\{(v_{11},C_1),(v_{12},C_1),…,(v_{nk},C_n)\}
\end{gather*}

Each chunk is effectively represented multiple times, once for each hypothetical prompt. This extends the coverage of how queries may be phrased and matched.

\begin{algorithm}[h]
\caption{HyPE Indexing Phase (offline)}
\label{alg:hype-index}
\begin{algorithmic}[1]
\Require Corpus $D=\{d_1,\dots,d_M\}$; chunker $\mathcal{C}$; generator LLM $G$; encoder $f$; prompts-per-chunk $k$
\Ensure  Vector index $E$ mapping embeddings to chunks
\State $E \gets \varnothing$
\ForAll{document $d \in D$}
  \State $C \gets \mathcal{C}(d)$ \Comment{split $d$ into coherent chunks}
  \ForAll{chunk $c \in C$}
    \State $Q \gets G\bigl(\textsc{GenerateQuestions}(c),\,k\bigr)$
    \ForAll{question $q \in Q$}
      \State $\mathbf v \gets f(q)$ \Comment{embed question}
      \State $E \gets E \cup \bigl\{(\mathbf v,\,c)\bigr\}$
    \EndFor
  \EndFor
\EndFor
\State \Return{$E$} 
\end{algorithmic}
\end{algorithm}

The retrieval process at runtime follows a standard approximate nearest-neighbor (ANN) search in the vector space. When a user query $q$ arrives, it is embedded into $q=f(q)$. The system then locates the nearest $v_{ij}$ vectors within $E$, and retrieves the associated chunks for final answer generation by an LLM. Although the pipeline remains structurally similar to a Naive RAG, the key difference is that HyPE matches questions against questions, rather than questions against chunk text.

At present HyPE treats every generated question in the set
\(Q_i=\{q_{i1},\ldots,q_{ik}\}\) with equal importance: each prompt is
embedded once and contributes a single vector to the index.
We do not yet attempt to decide which hypothetical questions are “better’’
or discard those that are less representative.  
Determining prompt quality is an open issue—and likely to be
domain-dependent.  
In settings where domain knowledge is available (e.g.\ biomedical
literature or legal texts) conditioning the LLM on that knowledge could
produce more accurate or stylistically appropriate questions, which in turn
should strengthen retrieval.  Investigating prompt scoring and
domain-specific generation therefore remains future work.

\begin{algorithm}[h]
\caption{HyPE Retrieval Phase (online)}
\label{alg:hype-retrieve}
\begin{algorithmic}[1]
\Require User query $q$; encoder $f$; index $E$; top-$k$
\Ensure  Relevant chunk set $\mathcal{R}$
\State $\mathbf v_q \gets f(q)$
\State $\mathcal{R} \gets \textsc{ANN\_Search}(E,\mathbf v_q,k)$ \Comment{$k$ nearest vectors}
\Return{$\mathcal{R}$}
\end{algorithmic}
\end{algorithm}

This question–question alignment increases the probability of finding the correct chunks for two main reasons. First, many embedding models exhibit style-based clustering~\cite{reimers2019sentence}. Texts of similar form (e.g., interrogative sentences) often lie closer in the vector space. As a result, a user’s real-world query naturally aligns more closely with the hypothetical prompts that share its interrogative style. Second, generating multiple hypothetical queries per chunk broadens the “semantic reach,” covering a wider range of possible question formulations. Even if a user query is phrased in a slightly different way, there is a higher chance that at least one of the chunk’s hypothetical questions closely corresponds to it.

Another advantage of HyPE lies in how it addresses the inherent chunking tradeoff in retrieval systems. If chunks are too large, their embeddings become less precise because they encode a mix of multiple concepts, making vector-based similarity less reliable~\cite{karpukhin2020dense}. Conversely, reducing chunk size improves embedding specificity but risks losing crucial surrounding context. HyPE mitigates this issue by ensuring that each stored vector represents a specific piece of information within a chunk, while retrieval still returns the entire chunk with its broader context. This allows the system to retain the benefits of detailed, fine-grained embeddings without sacrificing the context for accurate retrieval.

\begin{figure*}[ht]
    \centering
    \begin{minipage}{1\textwidth}
    \includegraphics[width=1\linewidth]{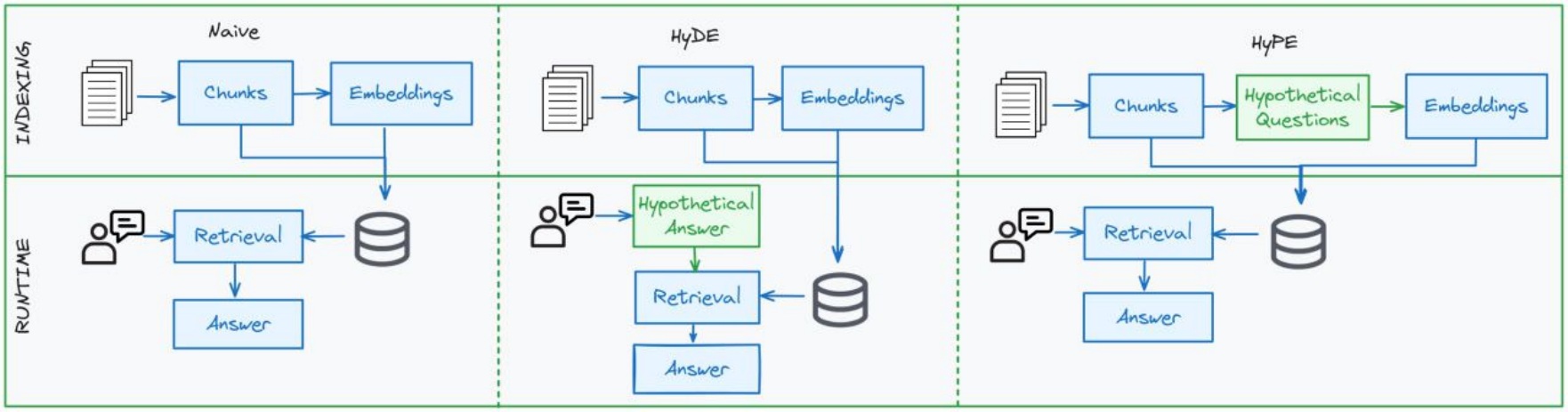}
    \caption{The image depicts the workflows of three retrieval-augmented generation (RAG) pipelines tested in the experiments: Naive RAG, HyDE, and HyPE. Blue components show parts of the pipeline that are the same across all setups and the green components show additional steps in the pipelines.}
    \label{fig:flows}
    \end{minipage}
\end{figure*}

HyPE invokes the language model once per chunk, prompting it to return a set of $m$ hypothetical questions in a single call. Consequently, a corpus consisting of $n$ chunks requires $n$ LLM calls during indexing, regardless of how many prompts are generated for each chunk. While this upfront cost can be substantial for very large datasets, it is paid only once and is strictly proportional to the dataset size. After the index is built, HyPE’s online path involves nothing more than standard vector search, incurring no additional LLM calls at query time and keeping serving latency and operating cost flat even as query volume grows.

\section{Experiments}
\label{experiment}
We evaluated three RAG pipelines to assess retrieval and generation performance as presented in Table~\ref{tab:my-table}. While a naïve retriever establishes how much can be achieved without any style-bridging, HyDE is the only published RAG component that explicitly targets the same “question-to-statement” gap as HyPE, albeit at inference time via synthetic-answer generation. Including HyDE therefore yields a good comparison and isolates the effect of moving hypothetical content creation from the query stage (HyDE) to the indexing stage (HyPE).

\begin{table}[h]
\centering
\begin{tabular}{|l|l|l|}
\hline
Retriever Pipeline  & Augmentation stage  & Context Space        \\ \hline
Naive RAG & /                   & prompt-to-document   \\ \hline
HyDE      & Inference           & document-to-document \\ \hline
HyPE      & Indexing            & prompt-to-prompt     \\ \hline
\end{tabular}
\caption{Key differences of compared pipelines.}
\label{tab:my-table}
\end{table}

\begin{figure*}[ht]
\centering
\begin{minipage}{1\textwidth}
  \centering
  \includegraphics[width=1\linewidth]{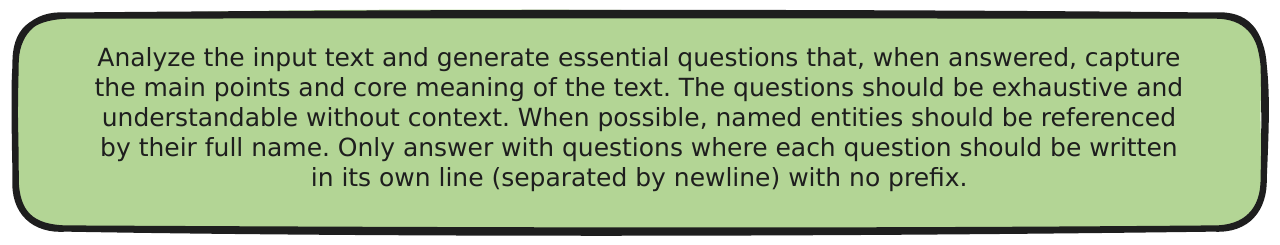}
  \caption{Prompt used to generate hypothetical prompts}
  \label{fig:prompt}
\end{minipage}%

\end{figure*}

For evaluating the RAG pipelines, we used the RAGChecker framework~\cite{ru2024ragchecker}, a comprehensive evaluation toolkit developed by Amazon Science for assessing both retrieval and generation performance in RAG systems. It provides structured metrics to analyse retrieval effectiveness through context precision, which measures how many retrieved passages are relevant, and claim recall, which assesses whether all necessary information is retrieved.

For generation, RAGChecker evaluates faithfulness, ensuring the generated text remains grounded in the retrieved passages, along with the hallucination rate, which identifies unsupported claims, and context utilization, which measures how effectively retrieved passages contribute to responses. Additionally, it assesses robustness through noise sensitivity, testing the system’s response to query variations, and self-knowledge, which quantifies the model’s ability to recognize when it lacks sufficient information. Additionally, it computes precision, recall, and F1 scores, providing an overall measure of retrieval and response accuracy~\cite{ru2024ragcheckerfinegrainedframeworkdiagnosing}. 

\subsection{Datasets}
\label{datasets}
We evaluated our approach on six datasets, chosen to test distinct aspects of RAG systems. \textit{MS MARCO}~\cite{ms_marco}, a large-scale question-answering benchmark, and \textit{Ragas-WikiQA}\footnote{https://huggingface.co/datasets/explodinggradients/ragas-wikiqa} evaluate general-purpose retrieval in real-world scenarios. \textit{RAG-dataset-12000}\footnote{\url{https://huggingface.co/datasets/neural-bridge/rag-dataset-12000}} and \textit{MultiHopRAG}~\cite{tang2024multihoprag} emphasize multi-hop reasoning, requiring systems to synthesize information across multiple documents. \textit{RAGBench}~\cite{friel2024ragbench} tests hybrid tasks demanding both precise retrieval and coherent generation. Finally, \textit{Single-Topic RAG} dataset\footnote{\url{https://www.kaggle.com/datasets/samuelmatsuoharris/single-topic-rag-evaluation-dataset}} focuses on narrow domains, assessing precision in specialized contexts. A concise summary of their key statistics is given in Table~\ref{tab:corpora}.

\begin{table*}[h]
\begin{minipage}{1\textwidth}
\caption{Descriptive statistics of the six datasets used in our study. For each dataset we list its thematic focus, the number of gold question–answer pairs, the total number of text chunks created by our preprocessing, and the average chunk length in tokens.}

\label{tab:corpora}
\centering
\renewcommand{\arraystretch}{1.1}
\begin{tabular}{|l|c|c|c|c|}
\hline
\textbf{Corpus} & \textbf{Domain / Focus} & \textbf{\# Q\&A pairs} & \textbf{\# Chunks} & \textbf{Avg.\ chunk len.\ (tokens)} \\ \hline
MS MARCO                 & Web search passages            & 82326 & 676193 & 82 \\ \hline
RAGBench                 & Mixed downstream tasks         & 73286       & 317563    & 173 \\ \hline
Ragas-WikiQA             & Wikipedia factoid QA           & 232     & 460    & 688 \\ \hline
RAG-dataset-12000        & General knowledge QA           & 9600    & 18321   & 378 \\ \hline
MultiHopRAG              & Multi-hop reasoning            & 2556     & 3101    & 443 \\ \hline
Single-Topic RAG         & Narrow domain articles         & 80     & 1324    & 467 \\ \hline
\end{tabular}

\end{minipage}
\end{table*}

All datasets already come pre-segmented into chunks, except for \textit{RAG-dataset-12000} that contains a single context block, and \textit{MultiHopRAG}, which contains multiple larger documents. For these two cases, we manually split the source text into segments of maximum 500 tokens, overlapping each segment by 50 tokens to preserve cross-boundary coherence. This approach, while straightforward, may not be optimal as the choice of chunking strategy can significantly impact retrieval effectiveness and quality of generation. Accordingly, we apply the same chunking procedure across all three pipelines for consistency in our experiments, but we note that different pipelines might benefit from tailored chunking strategies. We leave an in-depth exploration of chunk size, overlap, and other segmentation heuristics as a direction for future research.

\subsection{Evaluation metrics}
\label{evaluationmetrics}

For all pipelines, we tested retrieval depths $k\in\{1,3,5,10\}$. At $k=5$, we additionally compared cosine similarity and Euclidean distance functions to assess their impact on chunk relevance ranking. The embedding model chosen for all pipelines was \textit{bge-m3}~\cite{bge-m3} for dense vector representations. The generator LLM used was \textit{Mistral-NeMo}\footnote{\url{https://mistral.ai/news/mistral-nemo/}}.

We chose Mistral-NeMo because its openly released weights make the model easy for anyone to download and run, ensuring that every result in this paper can be reproduced without relying on a commercial API. During the time of testing, published benchmark scores place it in the top tier of 7-13 B open-source LLMs for instruction following and QA, so it provides competitive generation quality while remaining fully replicable.

\section{Results with analysis}
\label{analysis}
In Table \ref{tab:retrieval-table}, we compare the three retrieval methods across six datasets and varying numbers of retrieved chunks ($k$). Each cell reports context precision (how many of the retrieved chunks directly match the query’s needs) and claim recall (how many relevant pieces of information are captured). Overall, HyPE improves recall by about 16 percentage points and precision by about 20 percentage points compared to Naive RAG, on average. The difference can be even larger on specific datasets, such as \textit{Single-Topic RAG} at $k=1$ where HyPE surpasses Naive RAG by more than 40 percentage points in precision and at $k=10$ surpasses by 44.6 percentage points in recall. Although HyPE performs slightly below Naive RAG on \textit{MS MARCO} at $k=1$, it catches up or exceeds Naive RAG at deeper retrieval. For \textit{RAGBench} and \textit{Ragas-WikiQA}, HyPE also achieves strong gains, especially at lower $k$ values, indicating its ability to retrieve the correct chunks accurately. 

\begin{table*}[!t]
\centering
\small
\begin{tabular}{|l|ll|ll|ll|}
\hline
\rowcolor{white}
\multicolumn{1}{|c|}{\textbf{Dataset@k}} & 
\multicolumn{2}{c|}{\textbf{Naive}} & 
\multicolumn{2}{c|}{\textbf{HyDE}} & 
\multicolumn{2}{c|}{\textbf{HyPE}} \\ 
\hline
\multicolumn{1}{|l|}{\textit{}} & 
\textit{Precision} & \textit{Recall} & 
\textit{Precision} & \textit{Recall} & 
\textit{Precision} & \textit{Recall} \\
\hline
RAG-12000@1 & \gradient{56}55.8 & \gradient{35}34.7 & \gradient{55}55.1 & \gradient{33}33.3 & \gradient{83}\textbf{82.6} & \gradient{64}\textbf{63.6} \\
RAG-12000@3 & \gradient{37}36.5 & \gradient{44}44.2 & \gradient{37}36.6 & \gradient{43}42.8 & \gradient{73}\textbf{73.0} & \gradient{76}\textbf{76.2} \\
RAG-12000@5 & \gradient{31}31.3 & \gradient{49}49.0 & \gradient{32}32.0 & \gradient{48}47.9 & \gradient{67}\textbf{66.9} & \gradient{80}\textbf{80.1} \\
RAG-12000@10 & \gradient{26}26.4 & \gradient{56}56.1 & \gradient{27}27.4 & \gradient{56}55.6 & \gradient{56}\textbf{56.1} & \gradient{85}\textbf{84.6} \\

\hline
MS MARCO@1 & \gradient{74}\textbf{73.6} & \gradient{56}\textbf{56.2} & \gradient{70}69.8 & \gradient{52}52.4 & \gradient{69}68.7 & \gradient{50}50.2 \\
MS MARCO@3 & \gradient{70}\textbf{70.2} & \gradient{75}\textbf{74.5} & \gradient{67}67.2 & \gradient{70}70.0 & \gradient{67}66.9 & \gradient{70}70.2 \\
MS MARCO@5 & \gradient{68}\textbf{67.6} & \gradient{81}\textbf{80.6} & \gradient{66}65.5 & \gradient{77}76.8 & \gradient{66}65.6 & \gradient{77}77.3 \\
MS MARCO@10 & \gradient{62}61.5 & \gradient{86}\textbf{85.7} & \gradient{61}60.7 & \gradient{83}82.8 & \gradient{63}\textbf{62.5} & \gradient{84}84.0 \\

\hline
MultiHop@1 & \gradient{36}36.2 & \gradient{22}21.9 & \gradient{36}35.6 & \gradient{22}21.9 & \gradient{44}\textbf{43.7} & \gradient{27}\textbf{27.2} \\
MultiHop@3 & \gradient{33}32.7 & \gradient{35}35.3 & \gradient{32}31.9 & \gradient{34}34.4 & \gradient{42}\textbf{42.2} & \gradient{43}\textbf{43.2} \\
MultiHop@5 & \gradient{30}30.3 & \gradient{40}39.9 & \gradient{30}30.3 & \gradient{40}39.8 & \gradient{40}\textbf{40.1} & \gradient{51}\textbf{50.6} \\
MultiHop@10 & \gradient{26}25.8 & \gradient{46}46.2 & \gradient{27}26.8 & \gradient{48}47.6 & \gradient{38}\textbf{37.5} & \gradient{59}\textbf{59.2} \\

\hline
RAGBench@1 & \gradient{65}65.0 & \gradient{39}38.7 & \gradient{59}58.7 & \gradient{34}33.9 & \gradient{66}\textbf{65.7} & \gradient{40}\textbf{39.5} \\
RAGBench@3 & \gradient{63}62.8 & \gradient{54}54.2 & \gradient{57}56.5 & \gradient{49}48.5 & \gradient{63}\textbf{63.0} & \gradient{58}\textbf{58.0} \\
RAGBench@5 & \gradient{60}60.0 & \gradient{61}60.7 & \gradient{54}54.3 & \gradient{55}55.3 & \gradient{61}\textbf{60.7} & \gradient{66}\textbf{65.6} \\
RAGBench@10 & \gradient{55}55.1 & \gradient{69}68.8 & \gradient{50}49.9 & \gradient{64}63.8 & \gradient{57}\textbf{57.1} & \gradient{75}\textbf{74.6} \\

\hline
Single-Topic@1 & \gradient{29}28.7 & \gradient{16}15.6 & \gradient{23}22.5 & \gradient{9}8.5 & \gradient{69}\textbf{68.8} & \gradient{33}\textbf{32.9} \\
Single-Topic@3 & \gradient{28}27.5 & \gradient{23}22.9 & \gradient{21}20.8 & \gradient{18}18.2 & \gradient{65}\textbf{64.6} & \gradient{63}\textbf{63.0} \\
Single-Topic@5 & \gradient{25}25.2 & \gradient{27}26.8 & \gradient{25}24.5 & \gradient{29}28.6 & \gradient{65}\textbf{64.5} & \gradient{69}\textbf{69.0} \\
Single-Topic@10 & \gradient{21}21.2 & \gradient{37}36.8 & \gradient{20}19.6 & \gradient{36}36.4 & \gradient{63}\textbf{63.0} & \gradient{81}\textbf{81.4} \\

\hline
WikiQA@1 & \gradient{52}51.7 & \gradient{33}32.5 & \gradient{53}53.0 & \gradient{32}31.7 & \gradient{87}\textbf{86.6} & \gradient{61}\textbf{61.1} \\
WikiQA@3 & \gradient{47}47.4 & \gradient{57}57.1 & \gradient{49}48.7 & \gradient{57}56.7 & \gradient{84}\textbf{84.2} & \gradient{78}\textbf{77.9} \\
WikiQA@5 & \gradient{40}40.1 & \gradient{65}65.0 & \gradient{44}43.8 & \gradient{68}67.6 & \gradient{85}\textbf{84.7} & \gradient{86}\textbf{85.6} \\
WikiQA@10 & \gradient{33}33.4 & \gradient{79}79.1 & \gradient{38}38.2 & \gradient{79}78.9 & \gradient{82}\textbf{81.9} & \gradient{90}\textbf{90.4} \\
\hline
\end{tabular}
\caption{Performance comparison of retrieval methods across datasets using context precision and claim recall metrics (bigger is better) with varying numbers of retrieved context chunks (k). Gradient intensity reflects metric strength (lighter to darker green indicates lower to higher values), with bold entries highlighting the best-performing method for each configuration.}
\label{tab:retrieval-table}
\end{table*}

\begin{figure}[!ht]
\centering
\begin{minipage}{1\linewidth}
  \centering
  \includegraphics[width=1\linewidth]{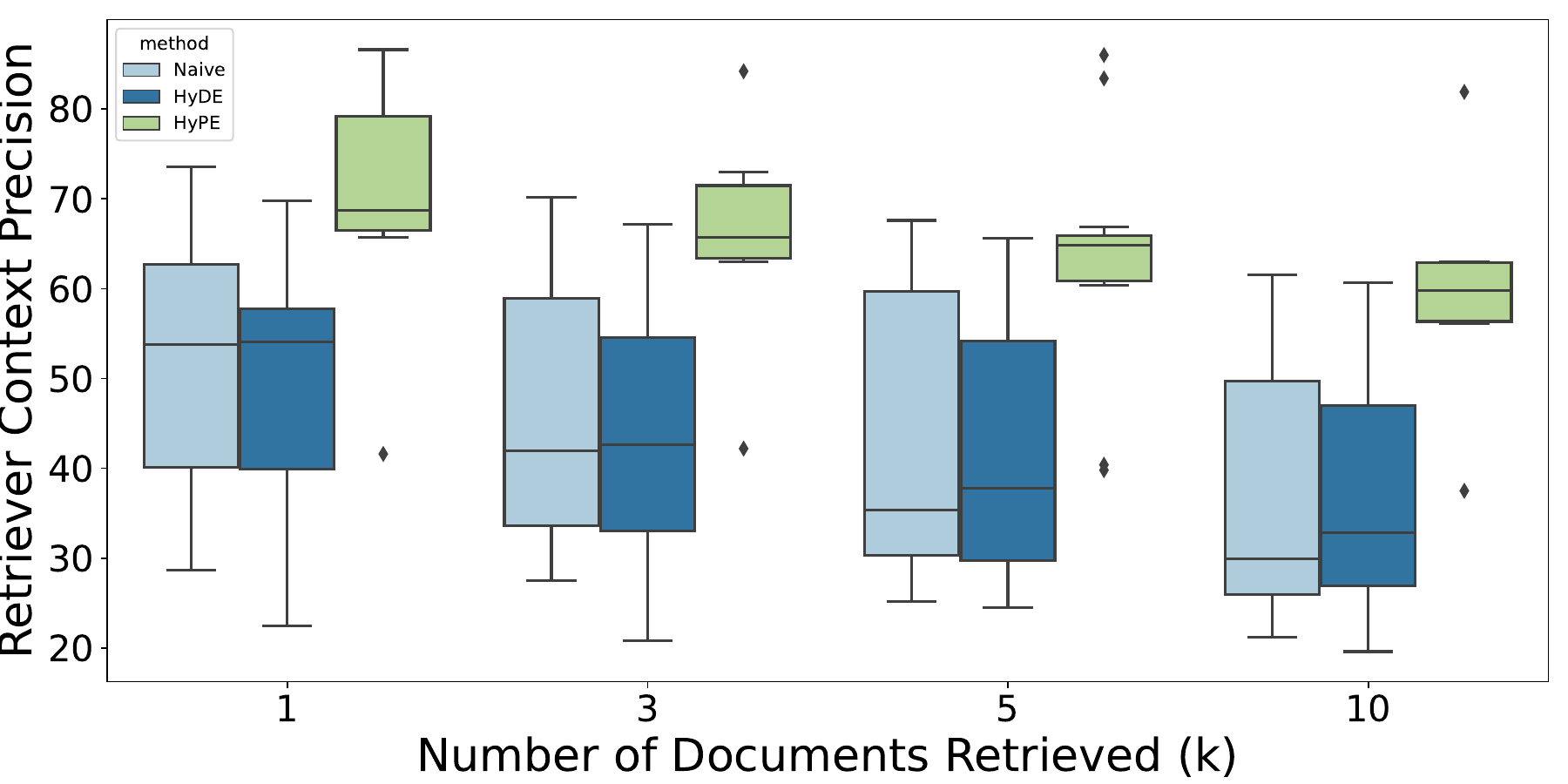}
  \caption{Box plot comparison of Retriever Context Precision across different numbers of documents retrieved (k) for three methods: Naive, HyDE, and HyPE. The plot illustrates the distribution and variability of precision scores for each method and retrieval depth.}
  \label{fig:percision}
\end{minipage}%
\end{figure}

Figure \ref{fig:percision} compares Retriever Context Precision across different numbers of documents retrieved (k) for the Naive, HyDE, and HyPE methods revealing significant improvement in precision, using HyPE. As the number of retrieved documents increases from 1 to 10, HyPE consistently demonstrates higher precision. Its precision is notably superior to both Naive and HyDE methods. This suggests that HyPE's approach of precomputing hypothetical questions during the indexing phase effectively aligns retrieved content with user queries, reducing the semantic mismatch often encountered in traditional methods. The narrower interquartile ranges for HyPE further indicate its consistency in retrieving relevant information across varying retrieval depths.

\begin{figure}[!t]
\centering
\begin{minipage}{1\linewidth}
  \centering
  \includegraphics[width=1\linewidth]{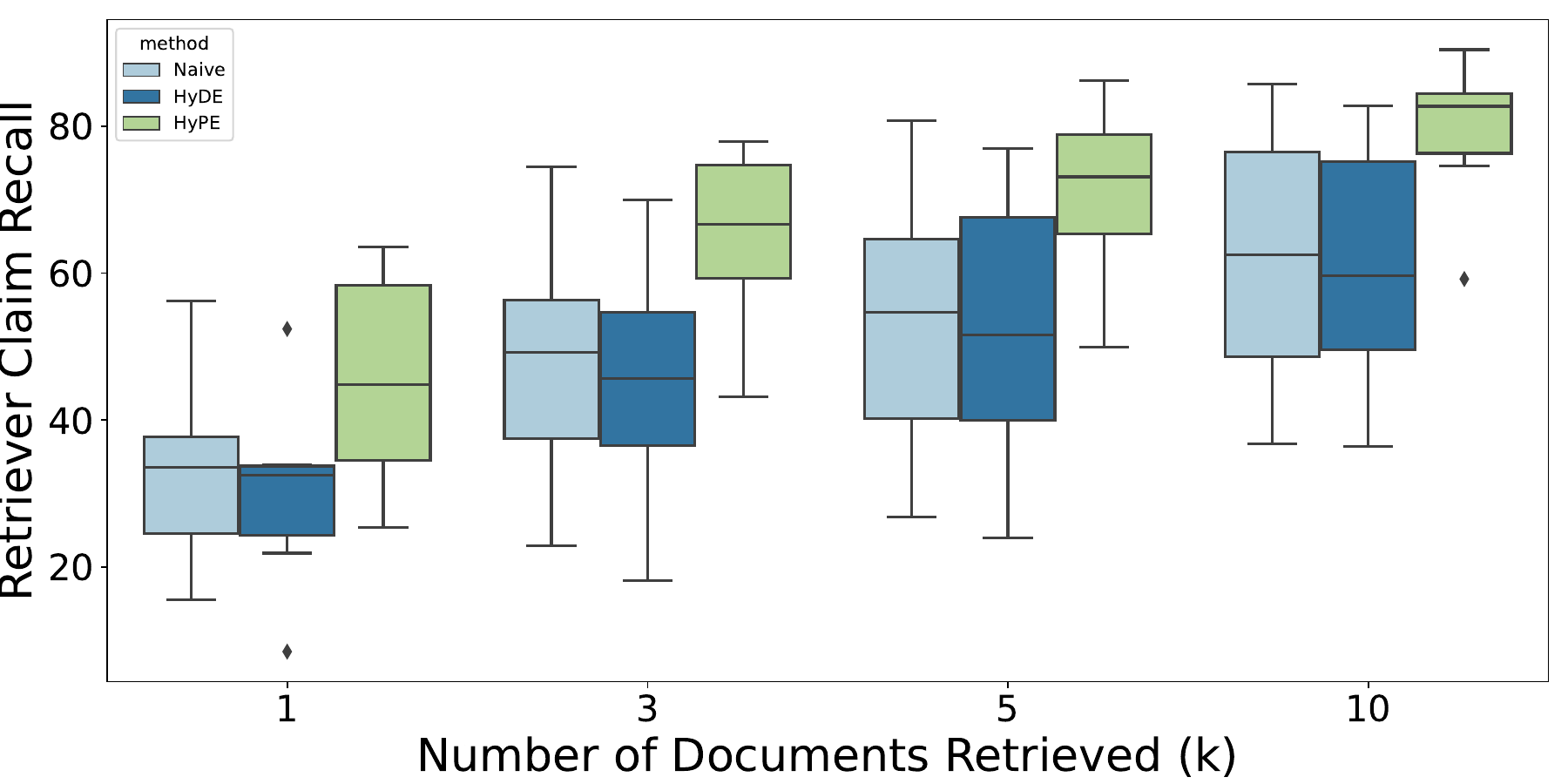}
  \caption{Box plot comparison of Retriever Claim Recall across different numbers of documents retrieved (k) for three methods: Naive, HyDE, and HyPE. The plot illustrates the distribution and variability of precision scores for each method and retrieval depth.}
  \label{fig:recall}
\end{minipage}%
\end{figure}

Additionally, the figure \ref{fig:recall} of claim recall complements these findings. Claim recall measures the proportion of relevant information successfully retrieved from the documents. The balanced performance in both metrics highlights HyPE's effectiveness in bridging the gap between user queries and relevant document content.

\begin{figure*}[h!]
\centering
\begin{minipage}{1\textwidth}
  \centering
  \includegraphics[width=1\linewidth]{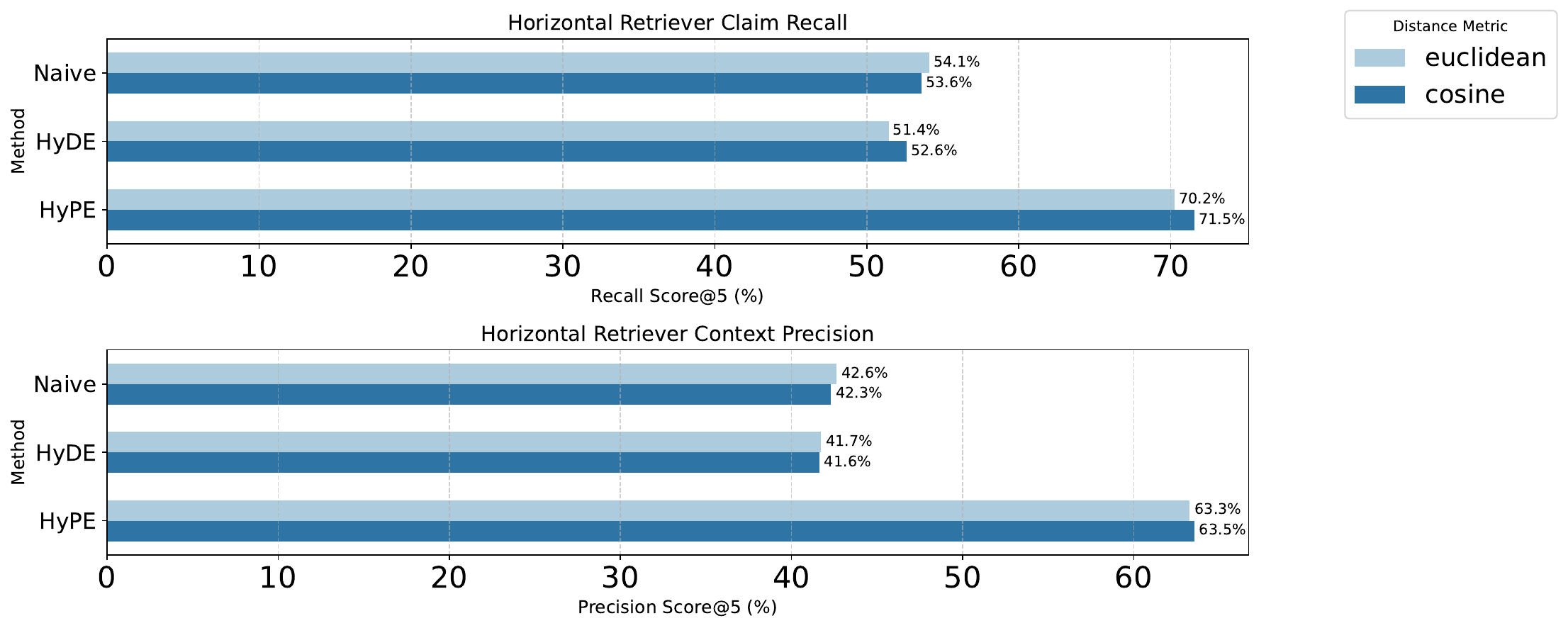}
  \caption{Comparison of Retriever Claim Recall and Context Precision across three retrieval methods using two distance metrics: Euclidean and Cosine. Performance measured at $k=5$.}
  \label{fig:distance}
\end{minipage}%
\end{figure*}

The comparison in Figure \ref{fig:distance} shows that for all three methods the choice between Euclidean and Cosine distance metrics does not significantly impact their effectiveness.  

\begin{figure*}[!h]
\centering
\begin{minipage}{1\textwidth}
  \centering
  \includegraphics[width=1\linewidth]{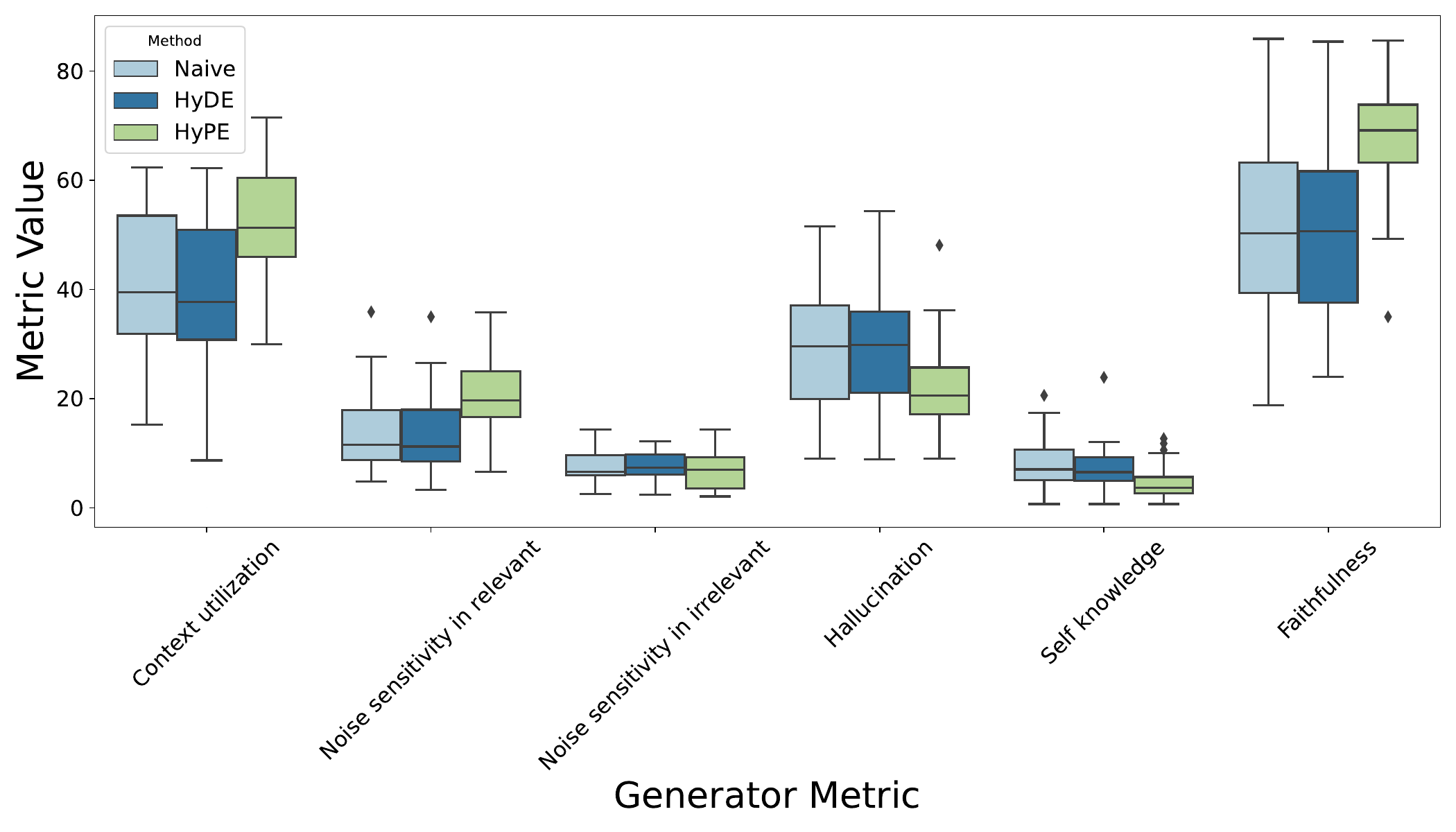}
  \caption{Comparative box plot analysis of various generator metrics when using one of the three retrieval methods, highlighting the performance differences among the methods in terms of their effectiveness and reliability in generating accurate and contextually relevant responses.}
  \label{fig:generator}
\end{minipage}%
\end{figure*}

In Figure \ref{fig:generator} we report generator metrics: context utilization, noise sensitivity, hallucination, self-knowledge, and faithfulness. These scores are properties of the generator language model, not of the retriever itself. Because HyPE intervenes only in the retrieval stage, we keep the generator LLM fixed and any other foundation model could be dropped in without changing the retrieval logic. Accordingly, shifts in these metrics across pipelines reflect how the quality of the retrieved context influences a given LLM’s behaviour, rather than inherent differences between language models.

\begin{table*}[h]
\caption{Aggregate performance across the six evaluation datasets
         (mean ± sd).  For metrics marked ↑, higher is better; for those
         marked ↓, lower is better.  Best value in each row is
         marked in bold.}
\label{tab:metric-summary}
\centering
\renewcommand{\arraystretch}{1.1}
\begin{tabular}{|l|c|c|c|}
\hline
\textbf{Metric} & \textbf{Naive} & \textbf{HyDE} & \textbf{HyPE} \\
\hline
Retriever claim recall ↑                 & 53.6 ± 19.0 & 52.6 ± 17.8 & \textbf{71.5 ± 12.5} \\
Retriever context precision ↑            & 42.3 ± 17.4 & 41.6 ± 15.9 & \textbf{63.5 ± 13.8} \\
\hline
Generator context utilisation ↑          & 40.0 ± 13.1 & 39.0 ± 15.7 & \textbf{53.5 ± 7.8}  \\
Generator faithfulness ↑                 & 52.2 ± 15.0 & 51.4 ± 14.8 & \textbf{69.3 ± 6.0}  \\
Generator hallucination ↓                & 26.0 ± 11.9 & 25.1 ± 11.4 & \textbf{19.9 ± 8.2}  \\
Noise sensitivity (irrelevant) ↓         &  7.5 ± 3.8  & \textbf{7.2 ± 2.7} & \textbf{7.2 ± 4.0} \\ 
Noise sensitivity (relevant) ↓           & \textbf{13.8 ± 7.8} & 14.2 ± 6.6 & 21.0 ± 4.4 \\
Self-knowledge ↓                         & 6.0 ± 3.4  & 5.4 ± 2.6 & \textbf{3.3 ± 1.4}  \\
\hline
Overall F1 ↑             & 27.9 ± 9.7  & 27.2 ± 9.6 & \textbf{37.6 ± 7.7} \\
Overall precision ↑                      & 35.2 ± 8.4  & 33.8 ± 7.5 & \textbf{42.6 ± 7.6} \\
Overall recall ↑                         & 37.9 ± 14.1 & 38.5 ± 13.9 & \textbf{50.4 ± 6.8} \\
\hline
\end{tabular}
\end{table*}

HyPE consistently achieves higher context utilization and faithfulness, indicating that its retrieval strategy provides more relevant and coherent supporting text for generation. However, the performance on noise sensitivity metrics presents a more nuanced picture. For 'Noise sensitivity in relevant contexts', HyPE registers a higher score compared to Naive RAG and HyDE. Within the evaluation framework \cite{ru2024ragchecker, ru2024ragcheckerfinegrainedframeworkdiagnosing}, this higher score signifies worse performance, suggesting the generator makes more errors when relevant retrieved documents are affected by noise. A potential explanation lies in HyPE's retrieval of multiple copies of the relevant chunks. Although beneficial for faithfulness through reinforcement, this very redundancy could increase the generator's susceptibility to errors since noise is repeated along with relevant information. Conversely, HyPE shows marginally better performance on 'Noise sensitivity in irrelevant contexts' (achieving a slightly lower score), indicating slightly improved handling of noise associated with irrelevant documents compared to the baseline methods.

HyPE also exhibits lower hallucination rates compared to Naive RAG and HyDE, reinforcing the idea that better-aligned retrieval reduces the likelihood of introducing incorrect or unsupported claims. Although these results are based on a single LLM, the trend is likely to generalize across different models, as improvements in retrieval typically translate to improved generation performance. However, the exact degree of impact may vary depending on the LLM’s retrieval dependence and sensitivity to context quality. The combination of improved context grounding, reduced hallucinations, and stronger response alignment highlights HyPE’s potential for enhancing the reliability of RAG systems.

\begin{figure*}[ht]
\centering
\begin{minipage}{1\textwidth}
  \centering
  \includegraphics[width=1\linewidth]{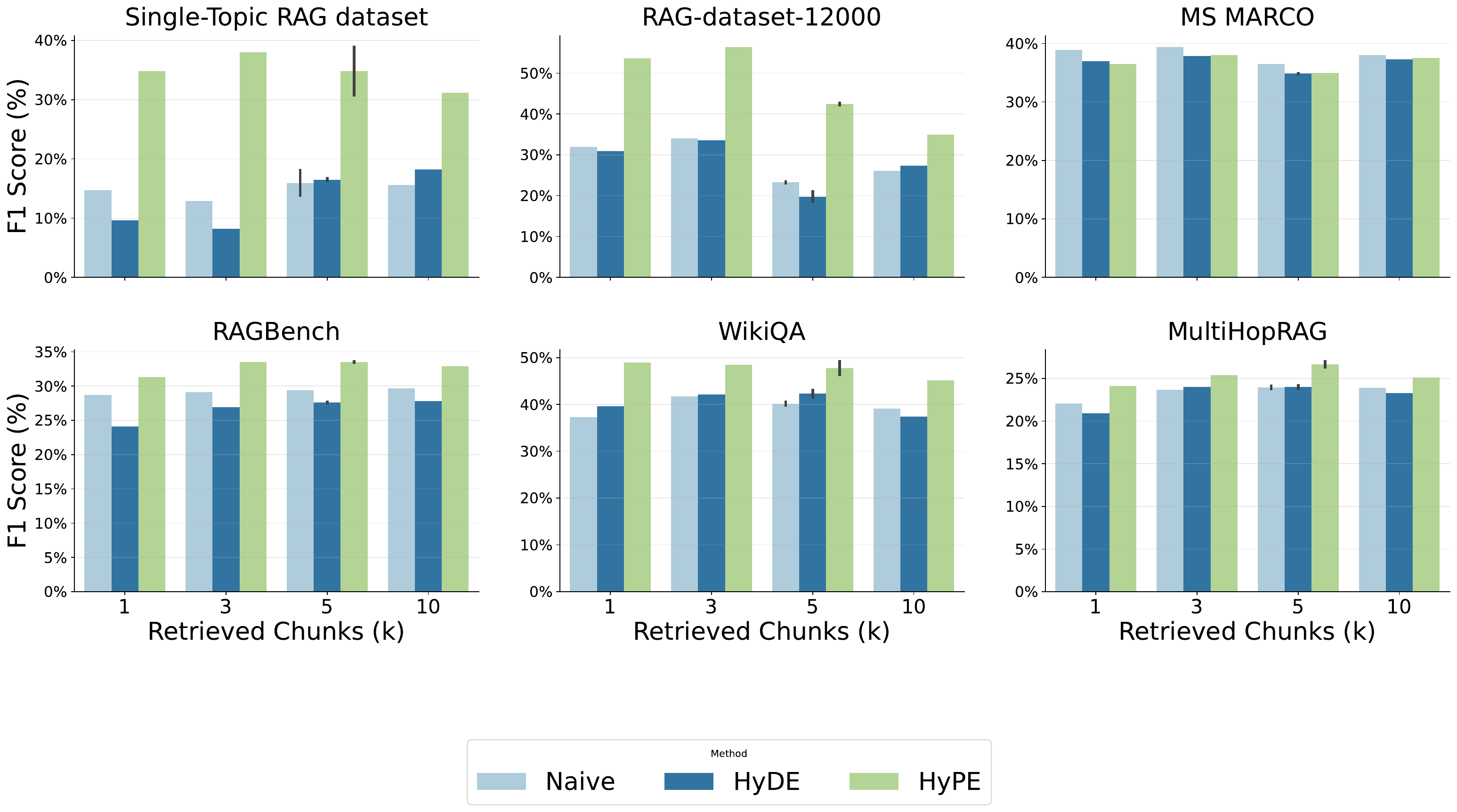}
  \caption{Bar chart comparison of F1 scores across six different datasets for three retrieval methods. Each subplot represents a dataset and shows the F1 scores for varying numbers of retrieved chunks (k = 1, 3, 5, 10). }
  \label{fig:f1}
\end{minipage}%
\end{figure*}

Figure~\ref{fig:f1} compares the overall F1 scores for retrievers and generators of the pipelines through the datasets. With the exception of the \textit{MS MARCO} dataset where all three methods show comparable performance, the F1 scores show, that HyPE consistently outperforms the other two methods, particularly in datasets like \textit{Single Topic RAG} and \textit{RAG-dataset-12000}, where the complexity and specificity of queries demand precise retrieval.

The MS MARCO benchmark stands out as an outlier in our results, exhibiting comparable retrieval efficacy across the Naive RAG, HyDE, and HyPE pipelines. This convergence can be attributed to several factors inherent to the dataset that limit the differential impact of HyPE's enhancements. Firstly, MS MARCO utilizes short, answer-centric passages (typically around 60 tokens), which facilitates strong baseline performance via direct query-passage matching, leaving less room for augmentation techniques to provide significant added value. Secondly, the high lexical overlap inherent between its web search-derived queries and corresponding passages substantially reduces the query-document style gap that HyPE aims to mitigate. Lastly, the high baseline retrieval scores observed in our experiments on MS MARCO, achieved even without augmentation using a powerful dense embedding model, suggest a performance saturation effect, where further architectural refinements struggle to yield significant marginal improvements. Consequently, while HyPE offers substantial gains on datasets characterized by longer documents and greater stylistic divergence between queries and context, its advantages are less pronounced given the specific properties of the MS MARCO passage retrieval task.

Given the small paired sample size and the absence of any firm evidence for normality, we adopt the distribution-free Wilcoxon signed-rank test with Holm–Bonferroni adjustment instead of a paired \(t\)-test.

\begin{table}[h]
\caption{Wilcoxon signed-rank tests (HyPE vs.\ each baseline).  Holm-adjusted \(p\)-values below 0.065 and Cliff’s \(|\delta|\!\ge\!0.56\) are bold.}
\label{tab:wilcoxon-main}
\centering
\renewcommand{\arraystretch}{1.05}
\begin{tabular}{lcc}
\hline
\textbf{Metric} & \textbf{HyPE vs Naive} & \textbf{HyPE vs HyDE} \\
\hline
Overall precision            & \textbf{0.094 / 0.56} & \textbf{0.063 / 0.72} \\
Overall recall               & 0.086 / 0.44          & \textbf{0.086 / 0.58} \\
Overall F1   & \textbf{0.063 / 0.56} & \textbf{0.063 / 0.61} \\
Retriever claim recall       & \textbf{0.063 / 0.61} & \textbf{0.063 / 0.67} \\
Retriever context precision  & \textbf{0.094 / 0.67} & \textbf{0.063 / 0.72} \\
Generator context utilisation & \textbf{0.063 / 0.67} & \textbf{0.063 / 0.61} \\
Noise sens.\ (relevant)     & \textbf{0.063 / 0.61} & \textbf{0.063 / 0.67} \\
Self-knowledge              & 0.086 / 0.64          & 0.086 / 0.64          \\
Hallucination              & 0.086 / 0.42          & 0.156 / 0.28          \\
Noise sens.\ (irrelevant)  & 1.000 / 0.03          & 1.000 / 0.11          \\
Faithfulness                & \textbf{0.063 / 0.67} & \textbf{0.063 / 0.72} \\
\hline
\multicolumn{3}{l}{\footnotesize Numbers are “adjusted \(p\) / \(|\delta|\)”.}
\end{tabular}
\end{table}

Across all eleven metrics, the Wilcoxon results in Table~\ref{tab:wilcoxon-main} indicate a consistent advantage for HyPE over both baselines. Nine metrics achieve an adjusted \(p<0.10\) in both comparisons, and five of those also satisfy the stricter \(p<0.065\) step that corresponds to the minimum attainable exact level. Effect sizes are uniformly medium to large: Cliff’s \(|\delta|\) ranges from 0.44 to 0.72, implying that a random sample from HyPE exceeds its baseline counterpart in roughly 70–85\% of the paired datasets. The only metric without a discernible difference is \textit{noise-sensitivity in irrelevant context} (\(p_{\text{adj}}=1.0,\;|\delta|<0.11\)), indicating that all methods degrade equally when faced with distractor passages. Taken together with the statistics in Table~\ref{tab:metric-summary}, these findings show that HyPE's improvements are both statistically reliable and practically meaningful.

Although HyPE is evaluated here as a stand-alone retriever, its design is orthogonal to other retrieval-side optimization approaches and therefore combinable with them. For example, the pre-computed question vectors can feed into query-decomposition modules (which break multi-hop questions into simpler sub-queries), query-expansion or rewriting steps (such as Doc2Query or RePlug), and even re-ranking or multi-vector fusion frameworks like BM25 or ColBERT. In these composite pipelines HyPE simply replaces the original passage vectors while leaving the higher-level orchestration intact, making it a drop-in upgrade rather than a competing subsystem. 

\section{Conclusion}
\label{conclusion}
The paper presents Hypothetical Prompt Embeddings (HyPE), a framework that pre-computes hypothetical prompts at indexing time to reshape retrieval in RAG pipelines into a prompt-to-prompt matching process. Our experimental findings show that HyPE surpasses both Naive RAG and HyDE on multiple datasets and metrics, with notable gains in precision and recall. By eliminating the need for query-time synthetic answer generation and instead relying on strategically generated questions offline, HyPE improves efficiency and provides stronger alignment between user queries and relevant content.

Although HyPE may not outperform every specialized RAG variant in all domains, it offers a flexible and modular upgrade to existing pipelines. Swapping in pre-computed question embeddings remains compatible with advances in chunking, re-ranking, multi-vector retrieval, and fine-tuning large language models. HyPE also integrates smoothly into agent-based systems, where prompt-level alignment can help specialized retrieval sub-agents handle distinct query types more effectively.

Looking ahead, combining HyPE with GraphRAG, which maps information from documents or chunks as graph nodes, may further enhance multi-hop reasoning and retrieval accuracy in complex scenarios. Such a hybrid approach could be especially valuable for building robust RAG systems.

Another direction for future research involves investigating the chunking tradeoff in the context of expanding LLM context windows. As language models evolve to accommodate larger input lengths, it becomes increasingly feasible to supply bigger chunks as prompts. However, larger chunks can dilute semantic specificity in their embeddings, resulting in less precise vector matching. This tension between maintaining detailed embeddings and preserving broader context may become more pronounced as context windows expand. Further testing of chunk size and indexing depth with HyPE would help clarify how embedding precision balances with retrieval breadth.

Finally, we plan to validate HyPE on multilingual RAG benchmarks to confirm that the question-question alignment holds across languages and writing systems.

Overall, HyPE demonstrates that shifting from question-to-document to question-to-question alignment leads to tangible gains in retrieval accuracy and cost-effectiveness. As RAG solutions continue to evolve, offline prompt generation strategies like HyPE can serve as a foundation for more efficient generation.


\bibliographystyle{IEEEtran}
\bibliography{refs}

@misc{ru2024ragcheckerfinegrainedframeworkdiagnosing,
      title={RAGChecker: A Fine-grained Framework for Diagnosing Retrieval-Augmented Generation}, 
      author={Dongyu Ru and Lin Qiu and Xiangkun Hu and Tianhang Zhang and Peng Shi and Shuaichen Chang and Jiayang Cheng and Cunxiang Wang and Shichao Sun and Huanyu Li and Zizhao Zhang and Binjie Wang and Jiarong Jiang and Tong He and Zhiguo Wang and Pengfei Liu and Yue Zhang and Zheng Zhang},
      year={2024},
      eprint={2408.08067},
      archivePrefix={arXiv},
      primaryClass={cs.CL},
      url={https://arxiv.org/abs/2408.08067}, 
}

@article{reimers2019sentence,
  title={Sentence-BERT: Sentence Embeddings using Siamese BERT-Networks},
  author={Reimers, N},
  journal={arXiv preprint arXiv:1908.10084},
  year={2019}
}

@misc{tang2024multihoprag,
      title={MultiHop-RAG: Benchmarking Retrieval-Augmented Generation for Multi-Hop Queries}, 
      author={Yixuan Tang and Yi Yang},
      year={2024},
      eprint={2401.15391},
      archivePrefix={arXiv},
      primaryClass={cs.CL}
}

@article{karpukhin2020dense,
  title={Dense passage retrieval for open-domain question answering},
  author={Karpukhin, Vladimir and O{\u{g}}uz, Barlas and Min, Sewon and Lewis, Patrick and Wu, Ledell and Edunov, Sergey and Chen, Danqi and Yih, Wen-tau},
  journal={arXiv preprint arXiv:2004.04906},
  year={2020}
}

@article{yasunaga2021qa,
  title={QA-GNN: Reasoning with language models and knowledge graphs for question answering},
  author={Yasunaga, Michihiro and Ren, Hongyu and Bosselut, Antoine and Liang, Percy and Leskovec, Jure},
  journal={arXiv preprint arXiv:2104.06378},
  year={2021}
}

@inproceedings{mishra2024searchd,
  title={Searchd-advanced retrieval with text generation using large language models and cross encoding re-ranking},
  author={Mishra, Pradumn and Mahakali, Aditya and Venkataraman, Prasanna Shrinivas},
  booktitle={2024 IEEE 20th International Conference on Automation Science and Engineering (CASE)},
  pages={975--980},
  year={2024},
  organization={IEEE}
}

@article{yepes2024financial,
  title={Financial report chunking for effective retrieval augmented generation},
  author={Yepes, Antonio Jimeno and You, Yao and Milczek, Jan and Laverde, Sebastian and Li, Renyu},
  journal={arXiv preprint arXiv:2402.05131},
  year={2024}
}

@misc{bge-m3,
      title={BGE M3-Embedding: Multi-Lingual, Multi-Functionality, Multi-Granularity Text Embeddings Through Self-Knowledge Distillation}, 
      author={Jianlv Chen and Shitao Xiao and Peitian Zhang and Kun Luo and Defu Lian and Zheng Liu},
      year={2024},
      eprint={2402.03216},
      archivePrefix={arXiv},
      primaryClass={cs.CL}
}

@article{ms_marco,
  title={Ms marco: A human-generated machine reading comprehension dataset},
  author={Nguyen, Tri and Rosenberg, Mir and Song, Xia and Gao, Jianfeng and Tiwary, Saurabh and Majumder, Rangan and Deng, Li},
  year={2016}
}

@article{ma2020zero,
  title={Zero-shot neural passage retrieval via domain-targeted synthetic question generation},
  author={Ma, Ji and Korotkov, Ivan and Yang, Yinfei and Hall, Keith and McDonald, Ryan},
  journal={arXiv preprint arXiv:2004.14503},
  year={2020}
}

@article{gao2022precise,
  title={Precise zero-shot dense retrieval without relevance labels},
  author={Gao, Luyu and Ma, Xueguang and Lin, Jimmy and Callan, Jamie},
  journal={arXiv preprint arXiv:2212.10496},
  year={2022}
}

@article{friel2024ragbench,
  title={Ragbench: Explainable benchmark for retrieval-augmented generation systems},
  author={Friel, Robert and Belyi, Masha and Sanyal, Atindriyo},
  journal={arXiv preprint arXiv:2407.11005},
  year={2024}
}

@article{ru2024ragchecker,
  title={Ragchecker: A fine-grained framework for diagnosing retrieval-augmented generation},
  author={Ru, Dongyu and Qiu, Lin and Hu, Xiangkun and Zhang, Tianhang and Shi, Peng and Chang, Shuaichen and Jiayang, Cheng and Wang, Cunxiang and Sun, Shichao and Li, Huanyu and others},
  journal={arXiv preprint arXiv:2408.08067},
  year={2024}
}

@article{lewis2020retrieval,
  title={Retrieval-augmented generation for knowledge-intensive nlp tasks},
  author={Lewis, Patrick and Perez, Ethan and Piktus, Aleksandra and Petroni, Fabio and Karpukhin, Vladimir and Goyal, Naman and K{\"u}ttler, Heinrich and Lewis, Mike and Yih, Wen-tau and Rockt{\"a}schel, Tim and others},
  journal={Advances in Neural Information Processing Systems},
  volume={33},
  pages={9459--9474},
  year={2020}
}

@article{izacard2022few,
  title={Few-shot learning with retrieval augmented language models},
  author={Izacard, Gautier and Lewis, Patrick and Lomeli, Maria and Hosseini, Lucas and Petroni, Fabio and Schick, Timo and Dwivedi-Yu, Jane and Joulin, Armand and Riedel, Sebastian and Grave, Edouard},
  journal={arXiv preprint arXiv:2208.03299},
  volume={1},
  number={2},
  pages={4},
  year={2022}
}

@article{nogueira2019doc2query,
  title={From doc2query to docTTTTTquery},
  author={Nogueira, Rodrigo and Lin, Jimmy and Epistemic, AI},
  journal={Online preprint},
  volume={6},
  number={2},
  year={2019}
}

@inproceedings{gospodinov2023doc2query,
  title={Doc2Query--: when less is more},
  author={Gospodinov, Mitko and MacAvaney, Sean and Macdonald, Craig},
  booktitle={European Conference on Information Retrieval},
  pages={414--422},
  year={2023},
  organization={Springer}
}

@article{nogueira2019document,
  title={Document expansion by query prediction},
  author={Nogueira, Rodrigo and Yang, Wei and Lin, Jimmy and Cho, Kyunghyun},
  journal={arXiv preprint arXiv:1904.08375},
  year={2019}
}

@article{furnas1987vocabulary,
  title={The vocabulary problem in human-system communication},
  author={Furnas, George W. and Landauer, Thomas K. and Gomez, Louis M. and Dumais, Susan T.},
  journal={Communications of the ACM},
  volume={30},
  number={11},
  pages={964--971},
  year={1987},
  publisher={ACM New York, NY, USA}
}

@article{robertson2009probabilistic,
  title={The probabilistic relevance framework: BM25 and beyond},
  author={Robertson, Stephen and Zaragoza, Hugo and others},
  journal={Foundations and Trends{\textregistered} in Information Retrieval},
  volume={3},
  number={4},
  pages={333--389},
  year={2009},
  publisher={Now Publishers, Inc.}
}

@inproceedings{xiong2021ance,
  title={Approximate Nearest Neighbor Negative Contrastive Learning for Dense Text Retrieval},
  author={Xiong, Lee and Wu, Chenyan and Yamada, Ikuya and Du, Zeyu and Lu, Xian and Yang, Jinfeng and Lin, Jimmy},
  booktitle={International Conference on Learning Representations (ICLR)},
  year={2021}
}

@article{eibich2024aragog,
  title={ARAGOG: A Large-Scale Evaluation of Retrieval-Enhanced Generation Methods},
  author={Eibich, Paul and Zhou, Nan and Rajani, Nazneen and Khashabi, Daniel and Roth, Dan and Durmus, Esin},
  journal={arXiv preprint arXiv:2404.01037},
  year={2024}
}

@article{gupta2024survey,
  title={A Comprehensive Survey of Retrieval-Augmented Generation: Evolution, Current Landscape and Future Directions},
  author={Gupta, Rishabh and others},
  journal={arXiv preprint arXiv:2410.12837},
  year={2024}
}

@article{cheng2025knowledge,
  title={A Survey on Knowledge-Oriented Retrieval-Augmented Generation},
  author={Cheng, Shitao and Lin, Bill Yuchen and Ren, Xiang},
  journal={arXiv preprint arXiv:2503.10677},
  year={2025}
}

\begin{IEEEbiography}[{\includegraphics[width=1in,height=1.25in,clip,keepaspectratio]{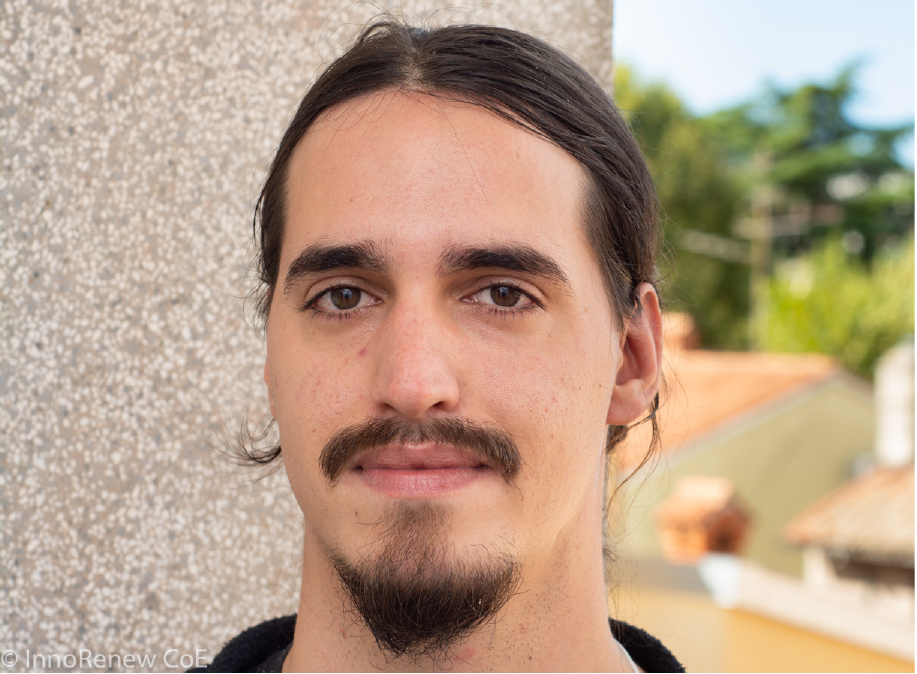}}]{\MakeUppercase{Domen Vake}} (M.S.) received the B.S. and M.S. degrees in Computer Science from University of Primorska, Slovenia. He is currently pursuing the Ph.D. degree in Computer Science at the Faculty of Mathematics, Natural Sciences and Information Technologies (FAMNIT) in the Department of Information Sciences and Technologies (DIST). Previously, he worked as a programmer at InnoRenew and now serves as an assistant researcher with affiliations both at InnoRenew and UP FAMNIT. His research interests include Artificial Intelligence with a larger focus on Large Language Models and their usage, as well as blockchain technology.
\end{IEEEbiography}

\begin{IEEEbiography}[{\includegraphics[width=1in,height=1.25in,clip,keepaspectratio]{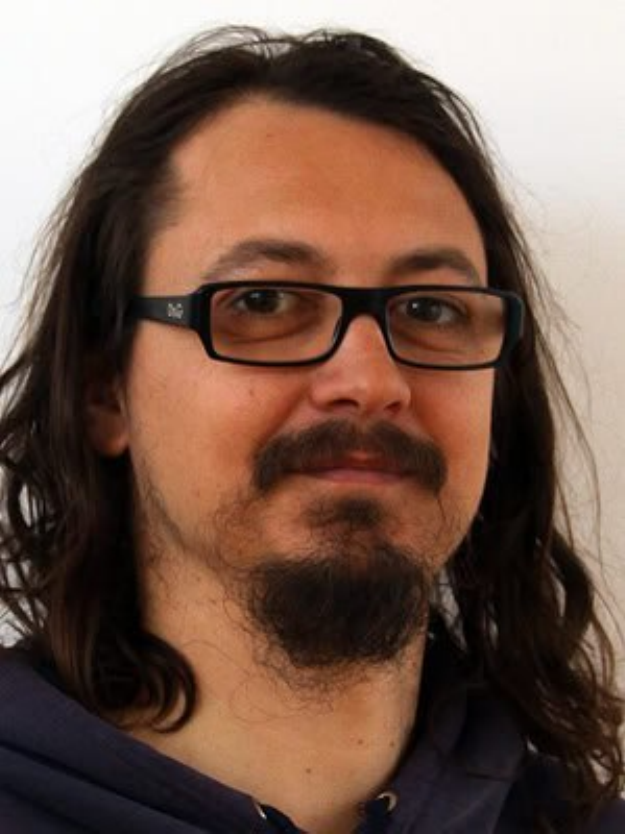}}]{\MakeUppercase{Jernej Vičič}} Associate professor and research associate at the University of Primorska and Research Centre of the Slovenian Academy of Sciences and Arts. His research focuses on artificial intelligence, natural language processing computational linguistics, and distributed systems. Jernej is also the head of the laboratory DLTLT at the UP FAMNIT. In 2023, Jernej Vičič received the Golden Plaque of the University of Primorska.
\end{IEEEbiography}

\begin{IEEEbiography}[{\includegraphics[width=1in,height=1.25in,clip,keepaspectratio]{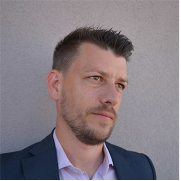}}]{\MakeUppercase{Aleksandar To\v{s}ić}} received his B.S. (2011), M.S. (2016), and Ph.D. (2022) degrees in Computer Science from the University of Primorska, Slovenia. He is currently an Assistant Professor at the University of Primorska and a Researcher at the InnoRenew CoE, Izola, Slovenia. He previously served as a Young Researcher at InnoRenew CoE and a Teaching Assistant at the University of Primorska. His research interests include distributed systems, privacy and security, sensors, and distributed ledger technologies. Dr. Tošić is a recipient of the Solemn Charter of the University of Primorska (2023) and the University Recognition for Academic and Research Achievements (2021). 
\end{IEEEbiography}

\EOD

\end{document}